
\magnification=1200
\hsize 15true cm \hoffset=0.5true cm
\vsize 23true cm
\baselineskip=15pt
\font\small=cmr8 scaled \magstep0
\font\grande=cmr10 scaled \magstep4
\font\medio=cmr10 scaled \magstep2
\outer\def\beginsection#1\par{\medbreak\bigskip
      \message{#1}\leftline{\bf#1}\nobreak\medskip\vskip-\parskip
      \noindent}

\def \pa {\partial}

\def \ti {\tilde}

\def \Da {\Delta}

\def \ga {\gamma}
\def \sg {\sigma}

\def \noi {\noindent}

\def\sqr#1#2{{\vcenter{\hrule height.#2pt\hbox{\vrule width.#2pt
height#1pt \kern#1pt\vrule width.#2pt}\hrule height.#2pt}}}

\def\lsim{\mathrel{\rlap{\lower4pt\hbox{\hskip1pt$\sim$}}
    \raise1pt\hbox{$<$}}}         
\def\gsim{\mathrel{\rlap{\lower4pt\hbox{\hskip1pt$\sim$}}
    \raise1pt\hbox{$>$}}}         

\nopagenumbers

\line{\hfil CERN-TH.6954/93}
\vskip 2.5 true cm
\centerline {\grande Quantum Squeezing and}
\bigskip
\centerline{{\grande Cosmological Entropy
Production}\footnote{$^{\dagger}$}{{\small Essay written
for the {\it 1993 Awards for Essays on Gravitation}
of the Gravity Research Foundation, and awarded with {\it Honorable
Mention}}.}}

\vskip 1 cm
\centerline{M. Gasperini \footnote{$^{\ast}$}
{{\small Permanent address: {\it Dipartimento di Fisica Teorica,
Via P.Giuria 1, 10125 Turin, Italy}}.}}
{\it
\centerline{Theory Division, CERN, Geneva, Switzerland} }
\centerline{and}
\centerline {M. Giovannini}
{\it
\centerline {Dipartimento di Fisica Teorica and INFN, Sezione di Torino,}
\centerline{Via P.Giuria 1, 10125 Turin, Italy} }

\vskip 1 cm
\centerline{\medio Abstract}
\noindent
The entropy growth in a cosmological process of pair production is completely
determined by the associated squeezing parameter, and is insensitive
to the number of particles in the initial state. The total produced entropy
may represent a significant fraction of the entropy stored today in the
cosmic black-body radiation, provided pair production originates from a
change in the background metric at a curvature scale of the Planck order.
\vskip 1.3 true cm
\centerline{To appear in {\bf Class. Quantum Grav.}}
\vskip 1 true cm
\noi
{CERN-TH.6954/93}\par
\noi
{July 1993}\par

\vfill\eject

\magnification=1200
\hsize 15true cm \hoffset=0.5true cm
\vsize 23true cm
\baselineskip=20pt
\font\small=cmr8 scaled \magstep0
\outer\def\beginsection#1\par{\medbreak\bigskip
      \message{#1}\leftline{\bf#1}\nobreak\medskip\vskip-\parskip
      \noindent}

\def \pa {\partial}

\def \ti {\tilde}

\def \Da {\Delta}

\def \ga {\gamma}
\def \sg {\sigma}
\def \Sig {\Sigma}

\def \noi {\noindent}

\def\sqr#1#2{{\vcenter{\hrule height.#2pt\hbox{\vrule width.#2pt
height#1pt \kern#1pt\vrule width.#2pt}\hrule height.#2pt}}}

\def\lsim{\mathrel{\rlap{\lower4pt\hbox{\hskip1pt$\sim$}}
    \raise1pt\hbox{$<$}}}         
\def\gsim{\mathrel{\rlap{\lower4pt\hbox{\hskip1pt$\sim$}}
    \raise1pt\hbox{$>$}}}         

\footline={\hss\rm\folio\hss}
\pageno=1

One of the greatest challenges of modern cosmic physics is that of
explaining
the large level of entropy observed on a cosmological scale. While it
 seems clear that inflationary kinematics has to play a fundamental
role in providing such an explanation,
the nature of the  (micro)physical mechanism that may have
 acted as a source of the cosmic entropy is, on the contrary, still
unclear.

A natural candidate for such a mechanism, well known even before the
advent of the inflationary models, is the production of particles by the
changing  background metric. The pair production from the vacuum
(otherwise stated, in a wave-mechanics language, the parametric
amplification of the background fluctuations) provides indeed a
natural cosmological arrow [1], which is not inverted even if the
expansion turns into a contraction, and which may thus be used to
define also an appropriate arrow of time [2]. The problem is, however,
that of quantifying in an unambiguous way, and in agreement with the
usual notion of entropy, the information loss associated with pair
production.

A possible solution to this problem comes from the observation that
the production of particles by an external gravitational field can be
conveniently represented in the squeezed state formalism [3]. Within
this context, one can define indeed two canonically conjugate quantum
variables (the so-called ``quadrature operators"[4]): one operator
has a variance that is ``squeezed" with respect to the vacuum, while
the variance of the other (the ``superfluctuant" one) is
correspondingly expanded.

Using this approach, we have recently proposed [5] a way to measure the loss
of information associated to the cosmological particle production,
in terms of the increased dispersion of the superfluctuant operator.
According to the standard (information-theoretic) definition of
non-equilibrium entropy, such a coarse graining approach
then provides an expression in which the entropy growth $\Da S_{k}$ is
completely determined, for each mode $k$, by the squeezing parameter
$r_{k}$ [5] :
$$
\Da S_{k}=2r_{k}.    \eqno (1)
$$
Our definition of entropy is consistent with analogous results,
recently obtained in the large squeezing limit, for the entropy of
the cosmological perturbations [6]. In our case, however, eq. (1) is
an exact result and can be applied also in the small squeezing regime.

By using eq. (1) we can estimate the entropy growth for a realistic case
in which particles (e.g. gravitons) are produced by the transition
between the inflationary era and the radiation-dominated era (occurring
at a given scale $H_{1}$).
We can assume, in particular, that their mode-number density has a
power law behaviour, characterized as usual by the spectral index $n$
(such that $k({d\rho/ dk})\simeq k^{4}\sinh^{2}r_{k}\sim k^{n-1}$).
In such a case the total produced entropy inside a co-moving horizon,
obtained by summing the entropy density (1) over all squeezed modes,
and measured in units of the usual black-body entropy $S_{\ga}$
associated to the electromagnetic CMB radiation, turns out to be [5]
$$
{\Da S\over S_{\ga}}\simeq\mid5-n\mid\left({H_{1}\over M_{P}}\right)
^{3/2} .              \eqno(2)
$$

This result is particularly interesting for a recently developed string
cosmology scenario [7], in which the spectral density of the cosmic
graviton background increases with frequency, in such a way that the
usual low-frequency bounds on the inflation scale are evaded, and
the maximum allowed value of
$H_{1}$ can be as large as
the Planck mass $M_{P}$ [8].
In such a context $H_{1}\sim M_{P}$ is indeed the natural transition scale,
so that the entropy stored in the relic graviton background could be
comparable to the observed large scale entropy of the black-body
radiation.

The expression (1) for the mode-entropy density,
however, was obtained in Ref. [5] for the ``minimum uncertainty"
squeezed vacuum state, i.e. under the assumption that the initial state,
whose fluctuation are amplified by the external gravitational
field, is the vacuum, or a ``displaced" vacuum (namely a coherent
state, in general). In this paper we shall show that
eq. (1) holds exactly, even if the initial state is any eigenstate of
the number operator or, more generally, any statistical mixture of
number states.

Consider indeed a pair production process described by a Bogoliubov
transformation connecting, for each mode $k$, the particle
($b,b^{\dagger}$) and antiparticle ($\ti b, \ti b^{\dagger}$)
annihilation and creation $ |in \rangle $ operators to the $|out \rangle
$ ones
($a,~a^{\dagger},~\ti a,~\ti a^{\dagger}$):
$$
a_{k}=\Sig_{k}b_{k}{\Sig^{\dagger}}_{k}~~~~~~~,~~~~~~~
\ti a_{-k}^{\dagger}=\Sig_{k}\ti b_{-k}^{\dagger}\Sig^{\dagger}_{k} .
\eqno(3)
$$
Here $\Sig_{k}$ is the two-mode squeezing operator [4],
$$
\Sig_{k}=\exp(z_{k}^{*}b_{k}\ti b_{-k}
- z_{k}b_{k}^{\dagger}\ti b_{-k}^{\dagger})~~~~~~,~~~~~
z_{k}=r_{k}e^{2i\vartheta_{k}} , \eqno(4)
$$
parametrized by the complex number $z_{k}$, which depends on the
dynamics of the external field leading to the process of pair creation.
In terms of the variables $x$ and $\ti x$, whose variance in the final
squeezed state is amplified with respect to their initial value, $a$
and $\ti a$ have the differential representation [5]
$$
a_{k}={i\over 2}e^{i\vartheta_{k}}\left[(\cosh r_{k}-\sinh r_{k})
(x-i\ti x) + (\cosh r_{k}+\sinh r_{k})(\pa_{x}-i\pa_{\ti x})\right]
$$
$$
\ti a_{-k}={i\over 2}e^{i\vartheta_{k}}\left[(\cosh r_{k}-\sinh r_{k})
(x+i\ti x) +(\cosh r_{k}+\sinh r_{k})(\pa_{x}+i\pa_{\ti x})\right]
\eqno(5)
$$
(the relative phase has been chosen with respect to $\vartheta_{k}$, in
such a way as to identify the $x$ and $\ti x$ operators with the
superfluctuant ones).

In the ($x, \ti x$)-space representation, the normalized squeezed
vacuum wave function  $\psi_{o,z_{k}}(x,\ti x)$ (such that
$a\psi_{o}=0=\ti a\psi_{o}$) is thus given by [5]
$$
\psi_{o,z_{k}}(x,\ti x)=\langle x\ti x|\Sig_{k}|00\rangle=
{\left(\sg\over \pi\right)}^{1\over 2}
e^{-{\sg_{k}\over 2}(x^{2}+\ti x^{2})}~~~~,~~~~\sg_{k}=e^{-2r_{k}} ,
\eqno(6)
$$
where $| 0\rangle$ is the $| in\rangle $ vacuum state, $b| 0\rangle =0
=\ti b| 0\rangle $. When $n$ pairs are already present in
the mode $k$ ($b_{k}^{\dagger}b_{k}| n_{k} n_{-k}\rangle =
n| n_{k} n_{-k}\rangle $),
however, the pair production leads to a two-mode squeezed number
state, whose wave function in this space has the explicit
representation
$$
\psi_{n,z_{k}}(x,\ti x)=~\langle x\ti x|\Sig_{k}| n_{k}n_{-k}\rangle~=~
\langle x\ti x| {{{(a_{k}^{\dagger}\ti a_{-k}^{\dagger})}^{n}}\over n!}
\Sig_{k}| 0\rangle~
$$
$$
= e^{-2in \vartheta_k}
\left({\sg_{k}\over\pi}\right)^{1\over 2}L_{n}
\left(\sg_{k}(x^2+\ti x^{2})\right)
e^{-{\sg_{k}\over 2}(x^{2}+\ti x^{2})}
$$
$$
=e^{in (\pi - 2 \vartheta_k )}
\left({\sg_{k}\over\pi}\right)^{1\over 2}
e^{-{\sg_{k}\over 2}(x^{2}+\ti x^{2})}
\sum_{m=0}^n {1\over m! (n-m)!}
H_{2m}(\sqrt \sg_k x) H_{2n-2m}(\sqrt \sg_k \ti x)
 \eqno(7)
$$
where $L_n$ and $H_n$ are the Laguerre and Hermite polynomials,
respectively.
It should be noted that, because of pair correlations, this wave function
cannot be simply factorized in terms of two decoupled squeezed
oscillators in an excited state, which are known to provide the usual
representation for the one-mode squeezed number wave function [9].
One can easily verify, however, that $x$ and $\ti x$ are superfluctuant
operators for the state (7), i.e.
$$
{(\Da x)^{2}_{n,z_{k}}\over {(\Da x)^{2}_{n}}}=
{(\Da \ti x)^{2}_{n,z_{k}}\over {(\Da\ti x)^{2}_{n}}}=e^{2r_{k}}>1 ,
\eqno(8)
$$
where
$$
(\Da x)_{n,z_{k}}^{2}=\langle n_{k}n_{-k}|\Sig_{k}^{\dagger}x^{2}\Sig_{k}
| n_{k}n_{-k}\rangle
$$
$$
(\Da x)^{2}_{n}=\lim_{z_{k}\rightarrow 0} (\Da x)^{2}_{n,z_{k}} \eqno(9)
$$
and the same for $\ti x$.

Suppose now that the initial state is a generic mixture of number
states,
$$
\overline\rho_{k}=\sum_{n} P_{n}(k)| n_{k}n_{-k}\rangle \langle
n_{k}n_{-k}|~~~,
{}~~~\sum_{n} P_{n}=1 . \eqno(10)
$$
The final state will correspond to the squeezed mixture:
$$
\overline\rho_{sk}
=\sum_{n} P_{n}(k)\Sig_{k}| n_{k}n_{-k}\rangle \langle
n_{k}n_{-k}| \Sig_{k}^{\dagger} .
 \eqno(11)
$$
Apart from the possible initial entropy, the pair production process is
accompanied by a loss of information corresponding to a larger
dispersion of $x$ and $\ti x$ around their mean values.
According to [5],
the reduced
density operator associated to such an increased uncertainty,
for the initial mixture (10), can be written
$$
\rho_{sk}
=\sum_{n}P_{n}(k)\int dx d\ti x \mid\psi_{n,z_{k}}(x,\ti x)\mid^{2}
| x\ti x \rangle \langle x \ti x| .    \eqno(12)
$$

By using the explicit form of the wave function (7), the net entropy
growth due to the pair production is thus
$$
\Da S_{k}= -Tr(\rho_{sk}\ln\rho_{sk})
+\left[Tr(\rho_{sk}\ln\rho_{sk})\right]_{\sg_{k}=1}=
-\sum_{n} P_{n}(k)\ln\sg_{k}=2r_{k}   \eqno(13)
$$
i.e. exactly the same as that obtained starting from the vacuum (eq. (1)).
This supports the internal consistency of our definition of entropy [5],
and the whole approach to the cosmological entropy based on the
squeezed-state formalism [5,6].

We conclude by noting that the result (13) has an important application
to the case of an initial thermal mixture.

The inflationary models
based on a temperature-dependent phase transition require indeed, as initial
condition, a thermal bath at a finite temperature higher than the
inflation scale. Such a non-zero temperature affects the spectral
energy density of the amplified perturbations, which may deviate from
the flat Harrison-Zeldovich form even in the case of pure exponential
inflation [10].
The result (13) shows instead that the produced entropy is not affected
by finite temperature corrections to the initial state. The entropy
depends only on the squeezing parameter (i.e. on the coefficients of
the Bogoliubov transformation, determined by the changing geometry),
and the expression (2) for the total entropy of a
given cosmic
background of relic particles is always
valid, irrespective of the initial field configuration.
\vfill\eject
\centerline {\bf References}
\item{1.}Hu B L and Pavon D, 1986 Phys.Lett.B180 329;

Hu B L and Kandrup H E, 1987 Phys.Rev.D35 1776;

Kandrup H E, 1988 Phys.Rev.D37 3505

\item{2.} Hawking S W, Laflamme R and Lyons G W, 1993 Phys.Rev.D47 5342

\item{3.}Grishchuk L P and Sidorov Y V, 1989 Class. Quantum Grav.6 L161;

1990 Phys.Rev.D42 3413;

\item{4.}Schumaker B L, 1986 Phys.Rep.135 317;

Grochmalicki J and Lewenstein M, 1991 Phys.Rep.208 189

\item{5.}Gasperini M and Giovannini M, 1993 Phys.Lett.B301 334

\item{6.}Brandenberger R, Mukhanov V and Prokopec T, 1992 Phys.Rev.Lett.69

3606; ``The entropy of the
gravitational field", Preprint Brown-HET-

849 (August 1992)

\item{7.}Gasperini M, Sanchez N and Veneziano G, 1991 Nucl.Phys.B364 365;

Veneziano G, 1991 Phys.Lett.B265 287;

Gasperini M and Veneziano G, 1992 Phys.Lett.B277 256;

Gasperini M and Veneziano G, 1992 ``Pre-big-bang in string cosmology",

Astroparticle Phys.1 (in press), CERN-TH.6572/92

\item{8.} Gasperini M and Giovannini M, 1992 Phys.Lett.B282 36;

1993 Phys.Rev.D47 1529

\item{9.} Kim M S, de Oliveira F.A.M. and Knight P L, 1990
``New frontiers in
quantum electrodynamics and quantum optics", ed. by A.O.Barut
(Plenum Press, New York) p.231

\item{10.}Gasperini M, Giovannini M and Veneziano G, 1993 ``Squeezed thermal
vacuum and the maximum scale for inflation", Phys.Rev.D48 (15 July 1993)

\end